\documentclass[prb,aps,twocolumn,draft,tightenlines,%
showpacs,floatfix]{revtex4}
\usepackage{psfig}
\usepackage{amssymb}
\usepackage{amsmath}

\begin{document}
\title{Intense terahertz laser fields on a two-dimensional electron 
gas with Rashba spin-orbit coupling}

\author{J. L. Cheng}
\author{M. W. Wu}
\thanks{Author to whom correspondence should be addressed}%
\email{mwwu@ustc.edu.cn.}
\affiliation{Hefei National Laboratory for Physical Sciences at
  Microscale and Department of Physics\footnote{Mailing Address.},
University of Science and Technology of China, Hefei,
  Anhui, 230026, China}

\date{\today}
\begin{abstract}
  The spin-dependent density of states and the density of
spin polarization of an InAs-based two-dimensional electron gas 
with the Rashba spin-orbit coupling under an intense terahertz
laser field are investigated by utilizing the Floquet states
to solve the time-dependent Schr\"odinger equation. 
 It is found that both densities are strongly affected by the 
terahertz laser field. Especially
a terahertz magnetic moment perpendicular to the external
terahertz laser field in the electron gas is induced.
This effect can be used to convert terahertz electric signals
into terahertz magnetic ones efficiently. 
\end{abstract}

\maketitle

Much attention has been devoted to the emerging field of
semiconductor spintronics recently.\cite{spintronics,das}
An active manipulation 
of the spin degrees of freedom in Zinc-blend semiconductors,
where the symmetry of the spin degrees of freedom is
broken due to the lack of  inversion center of the crystals,
is the central theme of this field. Many external
conditions, such as temperature and electric and magnetic fields,
can affect the spin coherence and have been discussed extensively
both experimentally and theoretically.\cite{das}
Among these conditions,  terahertz (THz) field, which 
can efficiently change the electron orbit momentums and significantly
modify the electron density of states (DOS), has been applied to 
spintronics rarely and only very recently,\cite{Johnsen,xu1}
although it has been extensively studied in 
spin-unrelated problems such as the dynamic Franz-Keyldysh
effect (DFKE) and optical 
sidebands.\cite{Yizhak,Cerne,Kono,Nordstrom,Phillips,zhang}
As an intense THz laser radiation affects the orbit degrees of freedom
efficiently, due to the spin-orbit coupling (SOC), it can also
effectively affect the spin degrees of freedom. In the present letter
we will show how a THz field can induce the spin polarization 
in a two-dimensional electron gas (2DEG) in
InAs quantum well by setting up
the Floquet states\cite{shirley} to solve  the time-dependent
Schr\"odinger equation. The effect discussed in the present letter
can be used to convert THz electric signals to THz magnetic ones.

We consider an InAs quantum well (QW) with its growth
direction along the $z$-axis. A uniform
THz field ${\mathbf E}_{THz}(t) = {\mathbf E}\cos(\Omega t)$ is applied
along the $x$-axis with the period
$T_0=\frac{2\pi}{\Omega}$. By using the Coulomb gauge, we write the
vector and scaler potentials into ${\mathbf A}(t) =
{\mathbf E}\sin(\Omega t)/\Omega$ and $\phi(t) = 0$ respectively. 
Then the total Hamiltonian is 
$H(t) = \frac{{\mathbf P}^2}{2 m^{\ast}} + H_{so}(\mathbf P)$
with ${\mathbf P} = -i\mbox{\boldmath$\nabla$\unboldmath} - 
e{\mathbf A}(t)$ (throughout the article we take $\hbar=1$) standing for the
electron momentum operator. $m^{\ast}$ is the effective mass of electron.
For InAs QW $H_{so}$ is dominated by the Rashba term
$H_{so}({\mathbf P, t})=\alpha[\sigma_x {\mathbf P}_y -
\sigma_y {\mathbf P}_x]$ 
which appears if the self-consistent potential within a QW
is asymmetric along the growth direction.\cite{Bychkov} 
\boldmath$\sigma$\unboldmath {} is the Pauli matrix and
$\alpha$ is the Rashba spin-orbit parameter which can be
as large as 4$\times 10^{-9}$\ eV\,cm.\cite{Grundler,Sato}
By taking the well width to be sufficiently small, one only needs to
consider the lowest subband.

With the help of the Floquet states, the solution of the 
Hamiltonian $H(t)$ can be written 
as, 
$\Psi_{{\mathbf k}, s}({\mathbf r},t)=\frac{1}{\sqrt{2\pi}}
e^{i{\mathbf k}\cdot{\mathbf r}}\Phi_{s}({\mathbf k}, t)$
with 
\begin{eqnarray}
  \Phi_{s}({\mathbf k}, t) &= &e^{-i\{(E_k + E_{em})t +
    r_0k_x[\cos(\Omega t) - 1] - \gamma\sin(2\Omega t)\}} \nonumber\\ 
  &&\times\sum_{n=-\infty}^{\infty}\phi_{n,
    s}({\mathbf k})e^{in\Omega
    t}e^{-iq_s(\mathbf k) t}\ .
\label{solution}
\end{eqnarray}
Here ${\mathbf k}=(k_x, k_y)$ stands for the 
electron momentum;  $s={\pm}$ represents two helix spin branches;
$E_{\mathbf k} =
{{\mathbf k}^2}/{2m^{\ast}}$ is the energy spectrum of electrons;
$E_{em}=\frac{e^2E^2}{4m^{\ast}\Omega^2}$ is an energy induced by the
radiation field due to the DFKE;
 $r_0={e E}/{m^{\ast}\Omega^2}$; $\gamma=E_{em}/(2\Omega)$.  
$\phi_{n,s}({\mathbf
    k})=(\phi_{n,s}^\sigma({\bf k}))
\equiv\genfrac{(}{)}{0pt}{}{\phi_{n,s}^{+1}({\mathbf
      k})}{\phi_{n,s}^{-1}(\mathbf k)}$ in Eq.\ (\ref{solution})
is the expansion coefficients of the  Floquet states\cite{shirley} 
of $H_{so}$, which gives the effect of the external THz field on the
 SOC with $\sigma=1$ ($-1$) representing spin-up $\uparrow$ (-down 
$\downarrow$) in the colinear (laboratory) coordinates (along $z$-axis). 
$q_s({\mathbf k})$ is the corresponding eigenvalue 
and can be determined by 
\begin{eqnarray}
&& [n\Omega-q_s({\bf k})] \phi_{n, s}^{\sigma} + \beta\sigma(\phi_{n-1,
  s}^{-\sigma}-\phi_{n+1, s}^{-\sigma})\nonumber\\
&&\mbox{}\hspace{1cm}+i\sigma\alpha k e^{-i\sigma\theta_{\mathbf
  k}}\phi_{n, s}^{-\sigma}=0\ ,
\label{quasi-energy}
\end{eqnarray}
in which $\theta_{\mathbf k}$ is the angle of the wave vector $\mathbf k$
and $\beta=\alpha m^{\ast} \Omega r_0/2$. All eigenvalues  can be written into
$n\Omega+q_s({\mathbf k})$ with $-\Omega/2<q_s({\mathbf
  k})<\Omega/2$. It is noted that
 $s=-$ branch can be determined by the $s=+$ one by 
$\phi^{\sigma}_{n,-}=-\sigma{\phi^{-\sigma\ast}_{-n,+}}$ and $q_{-}({\mathbf
  k})=-q_{+}({\mathbf k})$. It is seen from Eq.\ (\ref{quasi-energy})
that due to the SOC and in the presence of a THz field, two
spin branches (both helix spin branches  $s$ and colinear spin
branches $\sigma$) are {\em strongly correlated} to each other.
This correlation is a new feature due to the SOC and is beyond the
DFKE and the sideband effect.  In general, the
correlation is strong when the quantity $\lambda=\beta/(\Omega/2) \ge 1$. 

The retarded  Green functions in  momentum space at
zero temperature is 
\begin{eqnarray}
  G^{r}({\mathbf k}; t_1, t_2) =
  -i\theta(t_1-t_2)\sum_{s=\pm}\Phi_{s}({\mathbf
    k},t_1)\Phi_{s}^{\dagger}({\mathbf k}, t_2)
\end{eqnarray}
and the spectral function  $A=i(G^r - G^a)$ is therefore given by 
$A({\mathbf  k}; t_1, t_2) = \sum_{s=\pm1}\Phi_{s}({\mathbf
    k},t_1)\Phi_{s}^{\dagger}({\mathbf k}, t_2)$ which is a $2\times 
  2$ matrix in the spin space. After integrating over the momentum ${\bf
  k}$, one gets $\rho(t_1, t_2)=\frac{1}{(2\pi)^2}\int d{\mathbf k}
  A({\mathbf k};t_1, t_2)$. By letting $T=(t_1+t_2)/2$ and $t=t_1-t_2$ and
  transforming $t$ to the fourier space $\omega$,\cite{Jauho} one arrives at
\begin{widetext}
\begin{eqnarray}
  \rho_{\xi_1,\xi_2}(T, \omega)
 &=&\iint\limits_{-\infty}^{\ \ \ \infty}d{\mathbf
    k}\sum_{s=\pm}\hspace{0.3cm}\sum_{{{l_1,l_2,n,
    m}}=-\infty}^{\infty}R_{\xi_1,\xi_2}(s;  n, m; {\mathbf k})e^{i(n-m)\Omega T}
  J_{l_1}(2r_0{\mathbf  k}_x\sin(\Omega T))J_{l_2}(2\gamma
    \cos(2\Omega T))\nonumber\\ &&\times \delta(\omega-[E_{\mathbf
    k}+E_{em}-(l_1+2l_2+n+m)\Omega/2+q_s({\mathbf k})])\ , 
\label {total_result}
\end{eqnarray}
\end{widetext}
in which
$R_{\xi_1,\xi_2}(s; n, m; {\mathbf k}) = (\eta_{\xi_1}^{\dagger}\phi_{n,s}({\mathbf
  k}))({\phi_{m,s}}^{\dagger}({\mathbf k})\eta_{\xi_2})$ with
$\eta_{\xi}$ standing for the eigenfunction of $\sigma_z$ in colinear spin
system ($\xi=\sigma$) and the eigenfunction of $H(t=0)$, {\em i.e.},
$\eta_s=\frac{1}{\sqrt{2}}\genfrac{(}{)}{0pt}{}{1}{s\frac{ik_x-k_y}{k}}$, 
in helix spin system ($\xi=s$). It is
noted that in the absence of THz field, the off-diagonal term of $\rho_{\xi_1,\xi_2}$ 
vanishes in both the colinear and helix spin spaces (despite the fact that
the off-diagonal term of the spectral function $A$ in the colinear spin
space is not zero). However, introducing a THz field makes it non-zero in {\em both} systems.
A finite off-diagonal term of $\rho_{\xi_1,\xi_2}$ indicates the correlations
between the two spin branches and produces a spin polarization in the
2DEG. Similar as the diagonal term of $\rho$ is used to express
 the DOS, the off-diagonal term can be used to measure the density of the spin 
polarization (DOSP).

\begin{figure}[htbp]
\centerline{
  \psfig{figure=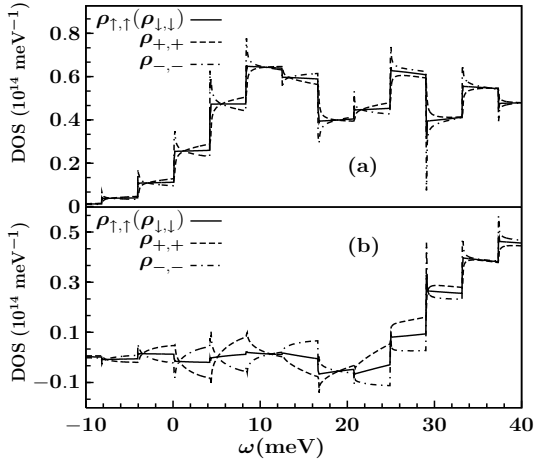,width=7cm,height=6cm}}
  \caption{The DOS at (a)  $T=0$
    and (b) $T=T_0/4$ with $\Omega=2\pi$
    THz and $E=6$ kV/cm. Solid
 curves: in the colinear spin space; Dashed and chain
 curves: in  helix spin space.}
  \label{fig:diagonal}
\end{figure}

From the symmetry of the Hamiltonian $H(t)$, one has $\Phi(k_x, -k_y, t)=
\sigma_y\Phi({\mathbf k},t)$ and $\Phi(-k_x, k_y,
t)=\Phi^{\ast}({\mathbf k}, -t)$. Therefore 
$\rho_{\xi_1,\xi_2}=\rho^{\ast}_{\xi_2,\xi_1}$ and 
$\rho_{\xi_1,\xi_2}(T_0 - T, \omega)=\rho^{\ast}_{\xi_1, \xi_2}(T, \omega)$.
Moreover in colinear spin space, one gets more meaningful properties:
$\rho_{\uparrow,\uparrow}=\rho_{\downarrow,\downarrow}$ and
 $\rho_{\uparrow,\downarrow}=-\rho^{\ast}_{\uparrow,\downarrow}$. 
Therefore the external THz field in $x$-axis can only excite 
the time-dependent spin polarization along the $y$-axis for the Rashba SOC.
Also from the fact that $\sum_\xi\rho_{\xi,\xi}$ does not change with the
spin space, one has $\rho_{\uparrow,\uparrow}=\rho_{\downarrow,\downarrow}
=\sum_s\rho_{s,s}/2$.

\begin{figure}[htbp]
\centerline{
  \psfig{figure=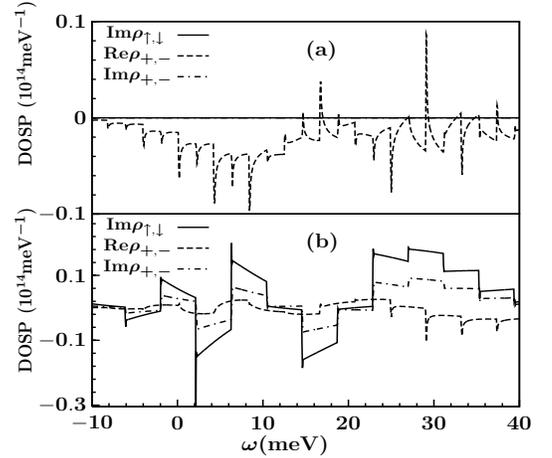,width=7cm,height=6cm}}
  \caption{The DOSP at (a) $T=0$ and (b) $T=
T_0/4$  with $\Omega=2\pi$\ THz and $E=6$\ kV/cm.
 Solid curves: The image part of the DOSP in the colinear spin space; Dashed
and chain curves: The real  and the image parts of the DOSP respectively
 in the helix spin space. It is noted that at $T=0$, 
Im$\rho_{+,-}=0$.}
  \label{fig:off-diagonal}
\end{figure}

We calculate the energy and time dependence of the
DOS and DOSP by numerically solving $\rho_{\xi_1,\xi_2}$ from Eq.\ (\ref{total_result}).
For InAs, $m^{\ast}=0.0239m_0$
 and $\alpha=3\times 10^{-11}$\ eV m,\cite{Grundler,Sato} with
 $m_0$ denoting the free electron mass. To show the general properties
 induced by the interaction between the THz field
 and the SOC, we choose $E\ge 1$\ kV/cm and $T_0=1$\ ps 
($\Omega=1$ THz) and 2.5\ ps (0.4\ tHz).  With these
 parameters, $\lambda$ is around 1, indicating a strong correlation between the
 two spin branches. The main results of our calculation are summarized 
in Figs.\ 1 to 3.

The DOS and DOSP are plotted in Figs.\ 1 and 2 in both spin 
spaces at $T=0$ and $T_0/4$. 
The DOS shows the effects from both the intense THz field and the SOC:
Both the sidebands shown as platforms in Fig.\ 1 and the blue-shifted  
main absorption edge\cite{Nordstrom} come from the intense THz 
fields. The Rashba SOC adds square root divergence peaks
at the left edge of each sideband. It also reduces the blue
shift $E_{em}$ by roughly $m^{\ast}\alpha^2/2$. It is
noted from the figure that although in the helix spin
system, the two spin branches are strongly separated by the THz field and the Rashba SOC,
they do not induce any spin polarization along the $z$-axis as 
in the laboratory (colinear) spin system $\rho_{\uparrow\uparrow}=\rho_{\downarrow\downarrow}$.
The most striking feature of the joint effect of the THz field and
the Rashba SOC comes from the {\em non-vanishing} off-diagonal terms of
$\rho$, which result in the DOSP as shown in Fig.\ 2 in both
spin spaces.

 \begin{figure}[htb]
\centerline{
  \psfig{file=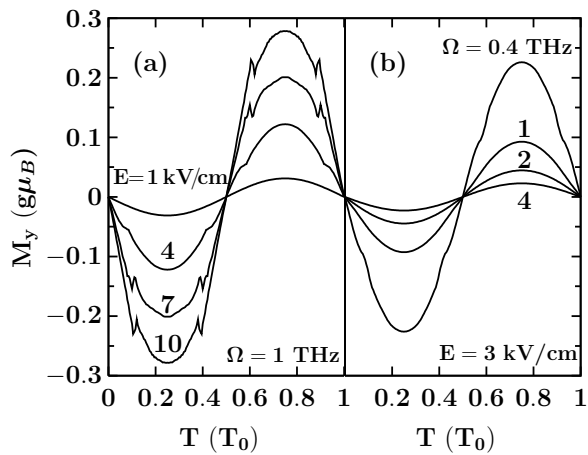,width=8cm,height=6cm}}
  \caption{The  average magnetic
    moment $M_y$  versus the time at 
    $E=1$, 4, 7 and 10 kV/cm for $\Omega=1$\ THz (a) at
$\Omega=0.4$, 1, 2 and 4\ THz for fixed $E=3$\ kV/cm (b).
The electron density is $N=10^{11}$ cm$^{-2}$.}
  \label{fig:fermi}
\end{figure}
 
By using $n_{\sigma} =
\frac{1}{2\pi}\int_{-\infty}^{E_f(T)}d\omega\rho_{\sigma,\sigma}(\omega,
T)$, one can determine the time-dependent Fermi energy 
$E_f(T)$. Then one obtains the average magnetic moment 
\begin{equation}
{\mathbf M}(T)=\left(0,
-\frac{g\mu_B}{n_{\uparrow}+n_{\downarrow}}
\int_{-\infty}^{E_f(T)}d\omega\ 
\mbox{Im}\rho_{\uparrow,\downarrow}(\omega, T),
0\right)\ .
\label{M}
\end{equation}
 Due to the THz field, the Fermi energy $E_f(T)$ and the average magnetic 
moment ${\bf M}(T)$ both oscillate with the period $T_0$.
In Fig.\ 3 ${\bf M}(T)$ is plotted as function of $T$ 
for fixed THz frequency $\Omega=1$\ THz and different electric
field strengths  $E=1$, 4, 7 and 10\ kV/cm (a) and
for fixed field strength $E=3$\ kV/cm and 
different frequencies $\Omega=0.4$, 1, 2
and 4\ THz (b)  with the electron density
$n_{\uparrow}=n_{\downarrow}=0.5\times 10^{11} \mbox{cm}^{-2}$. For
these parameters, for the 2DEG, the Fermi energy is about
10\ meV. It is seen from the figure that one obtains a THz magnetic 
signal which is induced by the THz electric one. The
strength of the magnetic momentum is controlled by the
electric field $E$ for fixed $\Omega$ and by the
THz frequency for fixed $E$. It is further pointed out that 
if  one substitutes  $E_f(T)$ with the
average value over the time period into Eq.\ (\ref{M}), 
one obtains the same oscillations 
in  ${\bf M}$-$T$ curves.

In conclusion we have proposed a scheme that can 
convert THz electric signals into THz magnetic ones
by calculating the DOS and DOSP of a 2DEG
with SOC and a uniform intense THz field.
The magnitude of the induced  magnetic signal is determined
by the amplitude of the electric field. This scheme 
has the potential 
to be applied to the magnetic resonance measurement non-magnetically.

This work was supported by the Natural Science Foundation of China
under Grant No. 90303012.  MWW was also supported by
the ``100 Person Project'' of Chinese Academy of
Sciences and the Natural Science Foundation of China under Grant No.
10247002. He would like to thank W. Xu for meaningful discussions.

\end{document}